\documentclass[aps,prb,twocolumn,showpacs,groupedaddress]{revtex4}
\usepackage{graphicx}

\begin{document}

\title{Pressure shift of T$_c$ in the system K$_x$Sr$_{1-x}$Fe$_2$As$_2$ (x=0.2, 0.4, 0.7): The analogy to the high-T$_c$ cuprate superconductors}
\author{Melissa Gooch$^1$, Bing Lv$^2$, Bernd Lorenz$^1$, Arnold M. Guloy$^2$, and Ching-Wu Chu$^{1,3,4}$}
\affiliation{$^{1}$TCSUH and Department of Physics, University of Houston, Houston, TX 77204, USA} \affiliation{$^{2}$TCSUH and Department of
Chemistry, University of Houston, Houston, TX 77204, USA} \affiliation{$^{3}$Lawrence Berkeley National Laboratory, 1 Cyclotron Road, Berkeley,
CA 94720, USA} \affiliation{$^{4}$Hong Kong University of Science and Technology, Hong Kong, China}
\date{\today }

\begin{abstract}
The systematic pressure shifts of T$_c$ were investigated in the whole phase diagram of the FeAs-based superconducting compound
K$_x$Sr$_{1-x}$Fe$_2$As$_2$. Different regions, arising from corresponding responses of T$_c$ to pressure (dT$_c$/dp$>$0, $\simeq$0, or $<$0),
can be clearly distinguished. This reveals an interesting similarity of the FeAs-superconductors and the high-T$_c$ cuprates. This behavior is a
manifestation of the layered structure of the FeAs compounds and the pressure-induced charge transfer between the (Fe$_2$As$_2$) and (K/Sr)
layers. The coexistence of superconductivity and spin-density wave behavior were also observed, and the pressure effects on the latter is
explored.
\end{abstract}

 \maketitle

The discovery of superconductivity in quaternary rare-earth transition-metal oxypnictides (ROTPn, where R=rare-earth, T=transition-metal and
Pn=pnictogen) with transition temperatures up to 26 K in F-substituted LaOFeAs and even higher T$_c$'s in compounds with smaller R-ions has
initiated immense excitement and stimulated renewed activities in high-T$_c$ superconductivity research.\cite{2,3,4,5,6,7} Only the copper oxide
superconductors are known so far to reach or exceed the high T$_c$ values of the FeAs-based compounds and it has been speculated that both
classes of superconducting materials are structurally and electronically related. The structural similarities are obvious, both compounds form
layered structures with active superconducting layers (Fe$_2$As$_2$ or CuO$_2$) separated by charge reservoir blocks providing control of the
nature and density of charge carriers. The Fe$_2$As$_2$ layers are extremely rigid allowing for significant variations of the charge reservoir
blocks which can even be a single plane of alkali or alkaline earth metals.\cite{8,9,9a} It is very important to show that other high-T$_c$
systems than the cuprate superconductors exist and to understand the common features of these compounds. This will facilitate the fundamental
understanding of high-T$_c$ superconductivity and possibly lead to the discovery of new materials with even higher critical temperatures.

To reveal the common properties of the FeAs superconductors and the cuprates and to verify their close relations we focus on the
T$_c$-dependence on the carrier density and the effect of pressure on T$_c$ for different doping levels. For the high-T$_c$ cuprates it is well
established that T$_c$ exhibits a universal doping dependence with a maximum at an optimal carrier density. Application of pressure transfers
charges from the reservoir to the CuO$_2$ planes and, accordingly, results in a positive, close to zero, and negative pressure coefficient of
T$_c$ for underdoped, optimally doped, and overdoped regions in the phase diagram, respectively.\cite{26} In the FeAs-based superconductors the
control of carriers is realized through the replacement of O with F (in ROFeAs),\cite{2,5,10,11,12,13,14} the control of oxygen
deficiency,\cite{15} or through the substitution of the alkaline earth (Ae) with a monovalent cation, K, Cs, etc. (in
AeFe$_2$As$_2$).\cite{16,17} While the maximum degree of F-substitution in ROFeAs is limited by the chemical stability and the region of highest
carrier concentration is difficult to achieve\cite{2,5,10,11,12,14} the K$_x$Sr$_{1-x}$Fe$_2$As$_2$ system was recently shown to be stable in
the whole doping range from x=0 (T$_c$=0) through optimal doping, $x_{opt}\simeq$0.45 (T$_c$=37 K), to x=1 (T$_c$=3.7 K).\cite{17} Similar phase
diagrams have been reported for Ba replacing Sr.\cite{18,19} This is the perfect system to be compared to the high-T$_c$ cuprates and to allow
for a thorough investigation of the pressure shift of T$_c$ in different regions of the phase diagram for low, optimal, and high carrier
densities.

Due to the complex Fermi surface of the FeAs-superconductors the effects of changes in carrier concentration on the electronic and
superconducting properties are not as clear as e.g. in the single-band high-T$_c$ cuprates. Band structure calculations reveal that the Fermi
surface includes electron as well as hole pockets.\cite{19a} For superconducting K$_x$Sr$_{1-x}$Fe$_2$As$_2$ near the optimal chemical
substitution ($x_{opt}$) the Hall coefficient was found to be positive hinting that the majority carriers are holes.\cite{19h} However the
negative Hall coefficient of the non-substituted Sr-122 indicates that electrons dominate the transport properties.\cite{19e} This
redistribution of states is expected since increasing $x$ reduces the number of electrons in the Fe$_2$As$_2$ layer and the majority carriers in
the K-doped compounds are holes. These results are consistent with recent ARPES measurements \cite{19i} and band structure
calculations.\cite{19a}

The application of high pressure and its effects on the superconducting T$_c$ were first investigated in the ROFeAs system. The increase of
T$_c$ with pressure initially observed in LaO$_{0.89}$F$_{0.11}$FeAs has raised the hope of achieving higher transition temperatures with
chemical or physical pressures in other R(O/F)FeAs superconductors.\cite{20} However, it was quickly shown that the pressure coefficient of
T$_c$ strongly depends on the doping level, and that the T$_c$ of SmO$_{1-x}$F$_x$FeAs ($x$=0.3, T$_c$=43 K) did actually decrease with
pressure.\cite{21} Related work on LaO$_{1-x}$F$_x$FeAs came to a similar conclusion. After the increase of T$_c$ at low pressure (p), for
compositions $x\leq$0.1, T$_c$ dropped precipitously with further increase of p.\cite{2} The negative pressure coefficient of R(O/F)FeAs
compounds with high T$_c$'s was also confirmed for LaO$_{0.9}$F$_{0.1}$FeAs, CeO$_{0.88}$F$_{0.12}$FeAs, and RO$_{0.85}$FeAs (R=Sm,
Nd).\cite{23,24,25} However, for a deeper understanding of the physics of the FeAs superconductors, and the suspected similarities to the
high-T$_c$ cuprates, a more systematic investigation of the effect of pressure for different carrier concentrations covering the complete phase
diagram is warranted.

We have therefore investigated the pressure dependence of T$_c$ in the phase diagram of K$_x$Sr$_{1-x}$Fe$_2$As$_2$ and compared our results
with those of the cuprate superconductors. We conclude that, similar to the high-T$_c$ cuprates, the main pressure effect can be explained by a
p-induced charge transfer between the reservoir and the active (Fe$_2$As$_2$) layer. In a limited range of $x$ the superconducting state below
T$_c$ was observed together with the spin-density-wave (SDW) state that is stable below a higher temperature T$_S$. The p-dependence of T$_S$ is
also studied.

Polycrystalline samples of K$_x$Sr$_{1-x}$Fe$_2$As$_2$ for $x$=0.2, $x$=0.4, and $x$=0.7 were prepared by high-temperature solid state
reactions, as previously described.\cite{17} The three samples are located in the phase diagram as shown in the inset of Fig. 2(a) (vertical
arrows). Hydrostatic pressure up to 18 kbar was generated in a piston-cylinder clamp cell using a Pb gauge for in situ pressure
measurements\cite{27} and liquid Fluorinert 70/77 as pressure transmitting medium. The ac magnetic susceptibility of two samples ($x$=0.4 and
0.7) was measured through a dual-coil inductance transformer wrapped to the sample's surface. For $x$=0.2 the resistivity was measured under
pressure since this also allows for extracting the p-dependence of T$_{S}$. The low-frequency (19 Hz) ac bridge LR700 (Linear Research) was
employed for inductance and resisitivity measurements.

\begin{figure}
\begin{center}
\includegraphics[angle=0, width=2.5 in]{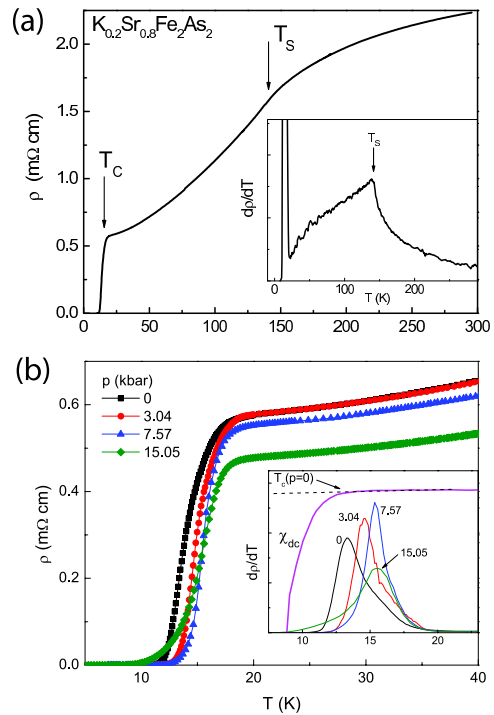}
\end{center}
\caption{(Color online) Resistivity of K$_{0.2}$Sr$_{0.8}$Fe$_2$As$_2$ vs. temperature at (a) ambient pressure and (b) high pressures. The
insets show the derivative with the anomalies at T$_S$ (a) and T$_c$ (b). The dashed line in the inset of (b) is the ambient pressure magnetic
susceptibility.}
\end{figure}

The $x$=0.2 sample deserves special attention since it shows two phase transitions easily detected in the resistivity data of Fig. 1(a). At
T$_S$=139 K the SDW transition is clearly resolved as a peak of the derivative d$\rho$/dT. At lower temperature, T$_c$=13 K, a superconducting
transition is defined by the drop of $\rho$(T) or another sharp peak of d$\rho$/dT. We use this peak position (inflection point of $\rho$(T)) to
define T$_c$ at ambient and high pressures. At ambient pressure the so determined critical temperature coincides with the onset of the
diamagnetic signal, as shown in the inset to Fig. 1(b). The region of coexistence of both transitions, SDW and superconductivity, extends from
$x\simeq$0.17 to $x\simeq$0.25 in the phase diagram shown in Fig. 2(a). This coexistence was also observed in
K$_x$Ba$_{1-x}$Fe$_2$As$_2$.\cite{18} The high-pressure resistivity data and their derivative are shown in Fig. 1(b). T$_c$ apparently increases
with p, but it saturates or approaches a maximum at the highest pressures obtained in this investigation. This is shown as the normalized shift,
T$_c$(p)/T$_c$(0), in Fig. 3 (top curve). Note that the resistivity drop broadens at the highest pressure, possibly due to pressure-induced
stress in the polycrystalline sample. The choice of the inflection point (which roughly corresponds to a 50 \% drop in $\rho$) is therefore a
reasonable criterion in assigning T$_c$. While T$_c$ increases with p, the SDW transition temperature T$_S$ (shown in the inset of Fig. 3)
decreases from 139 K (p=0) to 129 K at 15 kbar. The decrease of T$_S$ is consistent with a p-induced increase of the hole density in the
Fe$_2$As$_2$ layers. A similar suppression of the SDW state was observed in R(O/F)FeAs \cite{5} and K$_x$Ba$_{1-x}$Fe$_2$As$_2$.\cite{18}

\begin{figure}
\includegraphics[angle=0, width=2.5in]{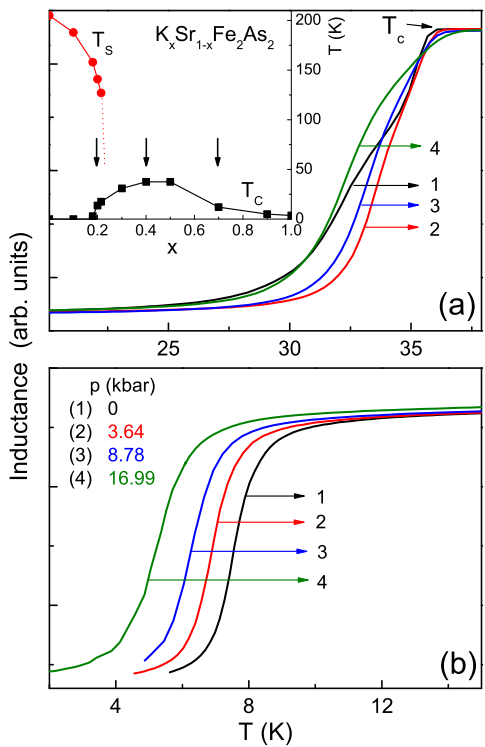}
\caption{(Color online) ac susceptibility of (a) K$_{0.4}$Sr$_{0.6}$Fe$_2$As$_2$ and (b) K$_{0.7}$Sr$_{0.3}$Fe$_2$As$_2$ near the
superconducting transitions at different pressures. The pressure values from 1 to 4 are labeled in Fig. 2b. The inset to Fig. 2a shows the phase
diagram of K$_x$Sr$_{1-x}$Fe$_2$As$_2$.}
\end{figure}

The ac susceptibility of a sample with maximum T$_c$ ($x$=0.4, T$_c$=37 K) is shown in Fig. 2(a). At zero pressure the diamagnetic response at
37 K is indicated by a sharp drop of the inductance but develops a shoulder-like feature at lower T ($\sim$32 K). This two-step magnetic
transition is characteristic of polycrystalline superconducting samples where intergrain coherence is established only at a lower temperature
than the intragrain superconducting transition. With increasing pressure the shoulder is diminished because of p-induced improvements of the
grain-grain contacts. At the highest pressure the inductance drop exhibits an overall broadening of the transition similar to the resistance
drop in the Sr-rich ($x$=0.2) sample. In addition, the change of T$_c$ with pressure is very small. In order to extract values of T$_c$ we
determined the onset temperature of the diamagnetic signal at different pressures. The normalized T$_c$(p) is plotted in Fig. 3 (center curve).
This data shows that the pressure shift of T$_c$ of this sample is negligibly small within the pressure range of this investigation. The
observed pressure behavior is similar to that of the cuprate superconductors near optimal doping.

In contrast to the two previously discussed samples, the ac susceptibility of the $x$=0.7 compound (largest hole concentration) exhibits a
significant decrease in the transition temperature of the diamagnetic signal (Fig. 2(b)). The inductance curves at different pressures are
nearly parallel, with T$_c$ decreasing rapidly and linearly with increasing p. The normalized pressure shifts of T$_c$ for all three samples are
shown in Fig. 3. From the data we conclude that, similar to the high-T$_c$ cuprate superconductors, the pressure effects on T$_c$ of the system
K$_x$Sr$_{1-x}$Fe$_2$As$_2$ depend strongly on carrier concentration levels. Moreover, considering the phase diagram T$_c$($x$) shown in the
inset to Fig. 2(a), we further conclude that the major effects of pressure is to increase the hole density in the Fe$_2$As$_2$ layers through
charge transfer to the charge reservoir block.

However, comparing the RO$_{1-x}$F$_x$FeAs and K$_x$Sr$_{1-x}$Fe$_2$As$_2$ superconductors, there are significant differences regarding the
pressure-induced charge control in the Fe$_2$As$_2$ layers. High pressure appears to increase the electron count in RO$_{1-x}$F$_x$FeAs
\cite{21} but it increases the hole density in K$_x$Sr$_{1-x}$Fe$_2$As$_2$. The two classes of FeAs superconducting compounds are distinguished
structurally by different stacking arrangements of the Fe$_2$As$_2$ layers along the c-axis. In the ThCr$_2$Si$_2$-type structure ("122"),
adjacent Fe$_2$As$_2$ layers are stacked in a way that two As atoms along the c-axis are related by a mirror plane resulting in closer As-As
distances and a doubling of the c-axis with respect to the unit cell of the "1111" compounds.\cite{17} In the ZrCuSiAs-type structure ("1111"),
the adjacent Fe$_2$As$_2$ layers are stacked "in phase" along the c-axis resulting in a smaller unit cell than in the "122" compounds. The
application of pressure on the "122" structure compresses the lattice, moves the As atoms to approach the K/Sr layers, and leads to shorter
interlayer distances. The shortening of the interlayer distance results in an increased hybridization of the Sr 5s/5p orbitals with the c-axis
component of the hybridized Fe/As orbitals and electrons are dispersed from the active Fe$_2$As$_2$ layer. This mechanism of pressure-induced
charge transfer is confirmed by recent band structure calculations.\cite{28} As a consequence the T$_c$ of the optimally doped sample is
expected to decrease with further increase in pressure, similar to the pressure dependence of T$_c$ derived from the onset of the resistivity
drop in LaO$_{0.89}$F$_{0.11}$FeAs.\cite{3} The small T$_c$ increase observed at low hydrostatic pressures in K$_{0.4}$Sr$_{0.6}$Fe$_2$As$_2$
can be understood as a response of a slightly underdoped sample. According to the phase diagram (inset, Fig. 2(a)) $x_{opt}$ lies between
$x$=0.4 and $x$=0.5.\cite{17}

\begin{figure}
\includegraphics[angle=0, width=2.5in]{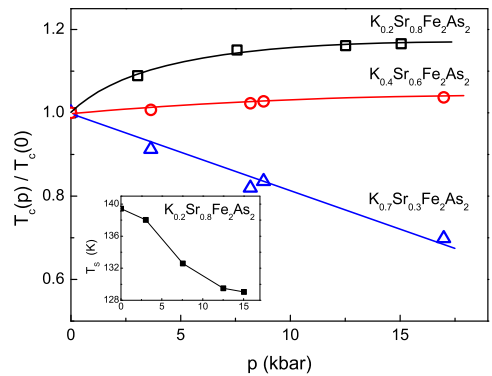}
\caption{(Color online) The shift T$_c$(p)/T$_c$(0) with pressure for K$_{0.2}$Sr$_{0.8}$Fe$_2$As$_2$, K$_{0.4}$Sr$_{0.6}$Fe$_2$As$_2$, and
K$_{0.7}$Sr$_{0.3}$Fe$_2$As$_2$. Inset: p-dependence of T$_S$ of K$_{0.2}$Sr$_{0.8}$Fe$_2$As$_2$.}
\end{figure}

The systematic investigation of the pressure effects on the superconducting T$_c$ of the system K$_x$Sr$_{1-x}$Fe$_2$As$_2$ has revealed
surprising similarities with the effects of pressure on the high-T$_c$ cuprate superconductors. This provides convincing evidence that the FeAs
superconductors are not only structurally but also electronically similar to the cuprates. From our results we also conclude that high pressure
increases the hole density in the Fe$_2$As$_2$ superconducting layers of the "122" compounds. This effect should also contribute to the
suppression of the SDW/structural transitions and the stabilization of superconductivity at high pressures in the non-superconducting compounds
AeFe$_2$As$_2$ (Ae=Ca, Sr, Ba).\cite{29,30,31} Accurate band structure calculations will be very useful in revealing the details of the Fermi
surface and, by varying the lattice parameters, in elucidating the effects of pressure on the charge distribution and carrier densities in the
Fe$_2$As$_2$ layers of the "122" family of superconducting FeAs-based compounds.

Note added: Recently we became aware of a pressure study of the related superconducting compound (K$_{0.45}$Ba$_{0.55}$Fe$_2$As$_2$) showing a
decrease of T$_c$ with pressure (Torikachvili et al., arXiv:0809.1080 [cond-mat.supr-con] (2008)).

\begin{acknowledgments}
This work is supported in part by the T.L.L. Temple Foundation, the
J.J. and R. Moores Endowment, the State of Texas through TCSUH, the
USAF Office of Scientific Research, and the LBNL through USDOE.
A.M.G. and B.L. acknowledge the support from the NSF (CHE-0616805)
and the R.A. Welch Foundation.
\end{acknowledgments}

\bibliographystyle{phpf}

\end{document}